# Enhanced 3D Myocardial Strain Estimation from Multi-View 2D CMR Imaging

Mohamed Abdelkhalek[1,3], Heba Aguib[1,3], Mohamed Moustafa[2], Khalil ElKhodary[1]

Mechanical Engineering Department, The American University in Cairo, Cairo 11835, Egypt.[1]

Computer Science Department, The American University in Cairo, Cairo, 11835 Egypt[2]

Biomedical Engineering & Innovation Lab, Aswan Heart Centre, Aswan, 81511, Egypt[3]

Correspondence should be addressed to Mohamed Abdelkhalek; mohamed-abdelkhalek@aucegypt.edu

## Abstract

In this paper, we propose an enhanced 3D myocardial strain estimation procedure, which combines complementary displacement information from multiple orientations of a single imaging modality (untagged CMR SSFP images). To estimate myocardial strain across the left ventricle, we register the sets of short-axis, four-chamber and two-chamber views via a 2D non-rigid registration algorithm implemented in a commercial software (Segment, Medviso). We then create a series of interpolating functions for the three orthogonal directions of motion and use them to deform a tetrahedral mesh representation of a patient-specific left ventricle. Additionally, we correct for overestimation of displacement by introducing a weighting scheme that is based on displacement along the long axis. The procedure was evaluated on the STACOM 2011 dataset containing CMR SSFP images for 16 healthy volunteers. We show increased accuracy in estimating the three strain components (radial, circumferential, longitudinal) compared to reported results in the challenge, for the imaging modality of interest (SSFP). Our peak strain estimates are also significantly closer to reported measurements from studies of a larger cohort in the literature and our own ground truth measurements using Segment Strain Analysis Module.  Our proposed procedure provides a relatively fast and simple method to improve 2D tracking results, with the added flexibility in either deforming a reconstructed mesh model from other image modalities or using the built-in CMR mesh reconstruction procedure. Our, proposed scheme presents a deforming patient-specific model of the left ventricle, using the commonest imaging modality , routinely administered in clinical settings, without requiring additional or specialized imaging protocols.

# Introduction

When we consider the multiplicity of mechanisms that underly the dynamics of a beating heart, such as the governing chemical, electrical, structural and fluidic processes, a need for sophisticated computational models becomes clear. Such models are typically designed to study only some aspect of a heart-beat, from the phenomena exhibited by blood flow within the chambers [1], to the electro-mechanics of the myocardium [2], to the evolving geometry and shape of the ventricular walls [3], to the associated deformation patterns of the ventricles [4]. Of general interest to such modelling efforts, is the translation of these governing principles to a computational domain, that is sufficiently representative of conditions in actual patients. This is necessary for the resulting models to be of clinical value and possibly assist in individualized therapy [5]. In particular, a study of the kinematics and deformation of the left ventricle (LV) on a patient-specific level is an example of computational modelling that has of recent gained traction in clinical research[5]). It can be divided into two approaches: tracking and analysis. Tracking falls within the fields of image processing and computer vision, since the underlying raw data that one starts with comes from a set of clinically acquired images in different modalities [6]. Some correspondence (image registration) then needs to be established between the images acquired at different times for a given region of interest (ROI) [7]. On the other hand, analysis includes methods from computational geometry and computational mechanics [8,9]. Quantities of clinical import that can be identified, and calculated for the LV, via both these approaches, include global metrics, such as chamber volume, wall thickness. And regional metrics, such as local displacement, strain and curvature (Fig. 1).

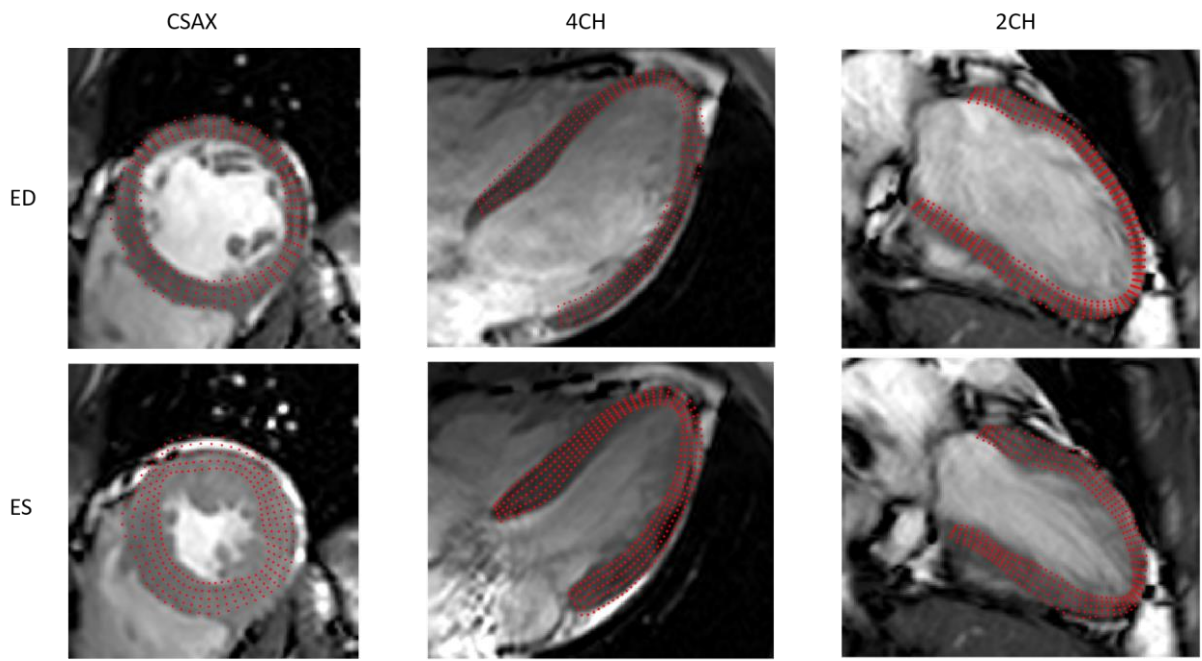

Figure 1. CMR SSFP images of the left ventricle in the different views at end-diastole ED and end-systole ES. Red points indicated tracked myocardium points using 2D registration method employed in Segment Medviso.

Strategies and methods are varied in terms of the choice of image acquisition, transformation modelling, similarity metrics, and post-registration error correction [7,8]. Historically, for instance, LV motion tracking was employed by direct methods of motion analysis within the frequency domain of the images [4,9]. Recently, there has been renewed clinical interest in LV motion tracking on untagged Cardiovascular Magnetic Resonance (CMR) images to help overcome the fading of tags over a cardiac cycle [10]. Moreover, interest in approaches that combine information from multiple imaging sources (multi-modality) has also grown [11]. In tandem, non-rigid registration methods that deal with cardiac deformation have also been widely adopted. These registration methods focus on elastic and optical flow methods that do not in general allow for motion tracking within the myocardium, but only along its borders [10,12]. Elastic methods seem preferable, since they add a smoothing penalty to the cost function in their optimization scheme, inspired by the bending energy of thin sheets of metal [13]. Furthermore, a temporal smoothing strategy could be introduced [14]. Their higher accuracy in comparison with other methods, e.g. optical flow, has thus been argued [12–14].

Table 1. Summary of previous work related to 3D myocardium motion tracking

| Group | PRINCIPLE | DIMENSIONALITY | PRE-PROCESSING | KEY FEATURES |
|---|---|---|---|---|
| MEVIS [15] | Quadrature filter-based elastic registration | 3D | Only in 3DUS, the images are down-sampled and then up-sampled afterwards | 2D registration in multiple directions on 3DTag, frequency based |
| IUCL [16] | Cross-correlation based on FFD | 3D | Mapping of 3DTAG and SSFP to common reference frame, valve tracking is used as an additional penalty term | Combined information of 3DTag and SSFP using a weighing scheme, uses valve tracking |
| UPF [17] | Temporal diffeomorphic FFD | 4D | 3D mask adjusted to cover myocardium over entire cycle | 4D registration using velocity instead of displacement |
| INRIA [18] | iLog Demons | 3D | Resampling of images, contrast enhancement and 3D mask | Incompressible, velocity field optimization |
| Segment [12–14] | Temporal coherent FFD | 2D | Contouring of CSAX, 4CH and 2CH at ED | FDA approved for clinical use, validated against speckle tracking US and variability was assessed in clinical setting |

As all approaches focused on assessing motion tracking, they do not elaborate on the pre-requirement of obtaining a geometrical representation of the LV for adaptation in real-life clinical setting and instead relied on the ground truth meshes provided by the benchmark challenge, which was obtained by manually deforming a template LV mesh acquired from CT image scans [4]. Therefore, for the purposes of this framework we included our own mesh reconstruction framework from SAX images to respond, to fact that in a clinical setting a template mesh may not be available, this also helps in creating LV shapes as close to real morphologies as possible. As a final remark, to the best of our knowledge Segment Medviso is one of the few FDA approved myocardial strain analysis commercial software that works on untagged CMR images without requiring special pre-processing steps, beyond those routinely administered for LV function assessment [14,19]. Hence, the motivation was to complement the software's clinical workflow with a) 3D strain interpolation and b) patient specific 3D mesh model of the LV, based on manual contouring of LV borders from CMR images.

# Materials and Methods

We begin by defining our problem as follows. We are given a sequence of images representing the heart over time, where 30 frames span about 1s of cardiac motion, imaging it as it contracts to eject blood, and then relaxes to refill, i.e. one cardiac cycle. The main image stack is of the short axis (SAX), which represents a top view of the heart chambers, and is composed of multiple 2D slices that trace the entirety of the LV from base to apex. Additionally, we have 2D image stacks of two- and four-chamber views that are orthogonal to the short axis (Fig. 1). The ROIs associated with the LV are three regions: the outer layer (epicardium), the inner layer (endocardium) and the volume between them (myocardium), which appear in 2D MRI images as per (Fig. 1). After assembling the image stack and loading them in a commercial software (Segment) we proceed to manually segment the ROIs for the first timeframe, which should always correspond to the maximal LV volume, known as the end-diastolic stage (ED). We then start the tracking process which follows the segmented ROI as it deforms over the cardiac cycle. We export the results of segmentation and tracking for coding on MATLAB and proceed to transform the tracked points into physical space, aligning the different views and up sampling their point clouds by means of spline interpolants. Next, we interpolate 3D displacement fields that combine motion data from the different views and reconstruct the surface and volume meshes from ED segmentation. This allows us to deform the new tetrahedral mesh representations of the LV (Video. 1), allowing for the calculation of strain distribution using finite element approximations on the deformation of each mesh element (Fig. 2)

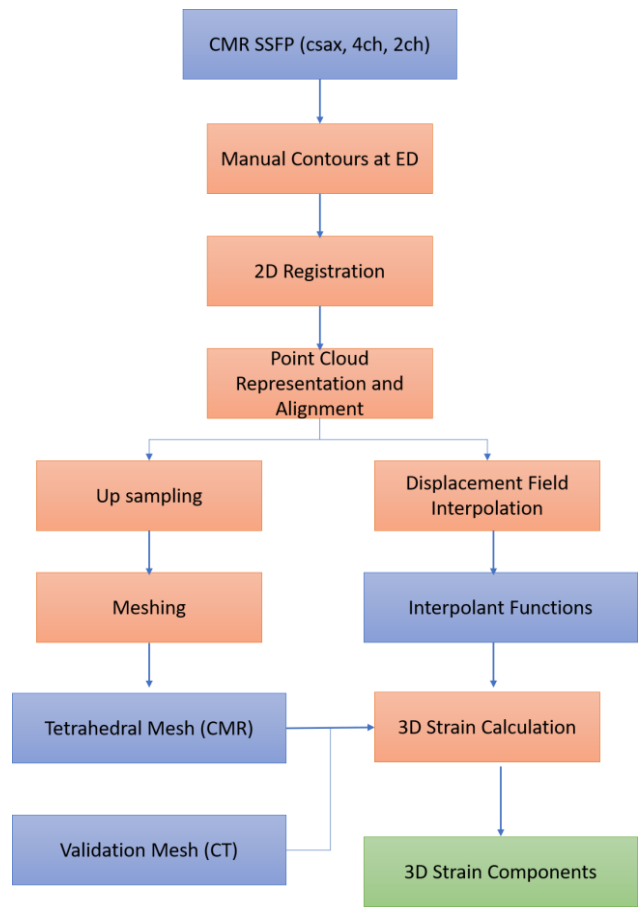

Figure 2. Block chart diagram showing the proposed method framework

**Segmentation and Initial Registration**

All pre-processing, which involves a consideration of DICOM metadata headers[6]) and manual segmentation, was performed on Segment. After loading and assembling the image stacks in the software, the next step is to manually segment the endo- and epicardial walls of the LV on the images using a graphical user interface tool provided by the same software (Fig. 1). This is a critical step that requires anatomical knowledge of the 3D morphology of the LV, its projections, and may be affected by image artefacts, noisy acquisitions, as well as observer experience. In our research we closely followed the segmentation guidelines recommended by the Task Force for Post Processing of the Society for Cardiovascular MR (SCMR) [19]. In Fig. 3A we show the process of contouring/segmentation of myocardial borders in the different slice levels.

Next, a tracking algorithm employed by the Segment Strain Analysis Module is used to estimate myocardial point displacements by using a non-rigid image registration method that has the following components. A transformation function using the B-Spline product transform, where control point displacements are estimated by an optimization scheme that minimizes an image similarity metric between pairs of consecutive images in the sequence [13,14]. Additionally, a regularization term that penalizes rough deformations is used in the optimization scheme. The similarity metric is treated simply as the sum of square differences SSD [13] between two consecutive image frames and the regularization term is based on the bending energy which can be derived from second-order derivatives of the cost function [20]. An important addition to this registration scheme, unlike other approaches, is its incorporation of time as an independent variable in the optimization process, which helps eliminate jagged or unphysical motion trajectories that are induced by treating inter-frame registrations independently [13]. This process is repeated for each view and for each 2D image in the stack. To counteract temporal misalignment, between long axis and short axis views, we reorder the image sequence such that all peak 2D strain time are aligned, and then we repeat the 2D registration. An important implication of the FFD approach employed by Segment is that we can also follow myocardial point displacements inside the tracked boundaries, unlike approaches based on optical flow methods [10]. The general form of a non-rigid elastic registration method based on FFD can be summarized as in [12,20] in (Eq. 1, 2, 3) as

$$C = S(\varphi) + \alpha R(\varphi) \quad (1)$$

$$S(\varphi) = \frac{1}{M} \sum_{i=0}^{M} \left( I_f(x) - I_m(T(x,\varphi)) \right)^2 \quad (2)$$

$$T(x) = x + u(x) \quad (3)$$

Where $C$ is the cost function and $S$ is the similarity measure (sum of square differences) between a fixed image $I_f$ and a moving image $I_m$ after applying a transformation $T$, which is defined as a displacement of some pixel/voxel located at $x$ defined by a deformation field $\varphi$. The $R$ term is the cost function that relates to smoothing/regularization with $\alpha$. The objective of the optimization is to find the optimal parameters of $T$ that minimize (Eq. 1). The choice of $R$, $T$, and $(x)$, as well as the strategy of obtaining the similarity measure $S$, are what determine the non-rigid registration algorithm. In this work, we complement the 2D tracking results provided by Segment Strain Analysis module which was used as is, detailed algorithm implementation details can be reviewed in[14].

**Point Cloud Construction and Alignment**

We next extract scattered displacement fields from the resultant tracked points by subtracting from the new positions at each frame (Fig. 3B) the original positions in the reference frame at ED (Fig. 3A). Upon exporting to MATLAB the sorted images, segmentation and tracking information, we need to perform a series of coordinate transformations that serve the dual purpose of aligning the different views which define the ROI, and facilitating the interpolation/extrapolation of the 3D displacement fields within a common coordinate system. This is accomplished by relying on the DICOM metadata information in each image stack, which provides the required information, i.e. position, orientation, scale and size, to allow a transformation of the images, segmentation and tracked points to the RL-AP-FH anatomical space [6]. We can then transform the resultant datasets to a global Cartesian coordinate system centred at the LV's centre of mass defined by averaging the nodal positions in the first time frame, with the major axes of the ROI aligned with X, Y, Z unit vectors respectively (Fig. 3B). The spatial resolution of the scattered points that are retrieved from segmentation, and their associated displacement fields, is in general deemed insufficient to fully reconstruct the LV. Hence, we up-sample the point clouds, relying on natural cubic splines [21,22] to construct spline cages that follow the ordered set of points that cover the ROI in its different spatial views.By stitching/joining endo and epi contours and then up-sampling points along these splines, we can generate higher quality point cloud representations of the LV (Fig. 3C), with the added ability to refine and smoothen our representation as needed, by means of predefined user variables. Generally, dense point clouds offer better surfaces and smoother final deformations.

**Meshing**

The up sampling described above satisfies the purpose of generatingan initial surface triangulation. However, a continued presence of gaps or holes in the 3D ROI can be an obstacle to creating a surface representation of the LV. We thus developed in-house a customized automatic method to stitch/join user segmentation data, and to create the desired closed surface (Fig. 3D), with the added feature of controlling our resolution/point-density via user defined variables. Our method takes advantage of experienced-user input during segmentation and allows true morphological features to emerge within our 3D representation. In particular, after a new set of points is retrieved (Fig. 3C), we proceed to generate an initial surface triangulation using the boundary fitting method MATLAB, which is similar to the convex hull method [23], but further allows the surface to shrink towards the interior of the hull to envelope all the points [23]. Subsequently, the surface is resampled, and a volume mesh is created using the iso2mesh utilities [24] which are based on Delaunay tetrahedralization of a closed surface. With our new set of nodes and their tetrahedral connectivity list, we obtain a complete volumetric representation of the LV (Fig. 3E) It is important to note that in case a surface mesh is already available, for example from an external segmentation process, one only needs to align the surface mesh to the CMR image set and construct a volume mesh from the surface representation using the same utilities in the iso2mesh toolbox.

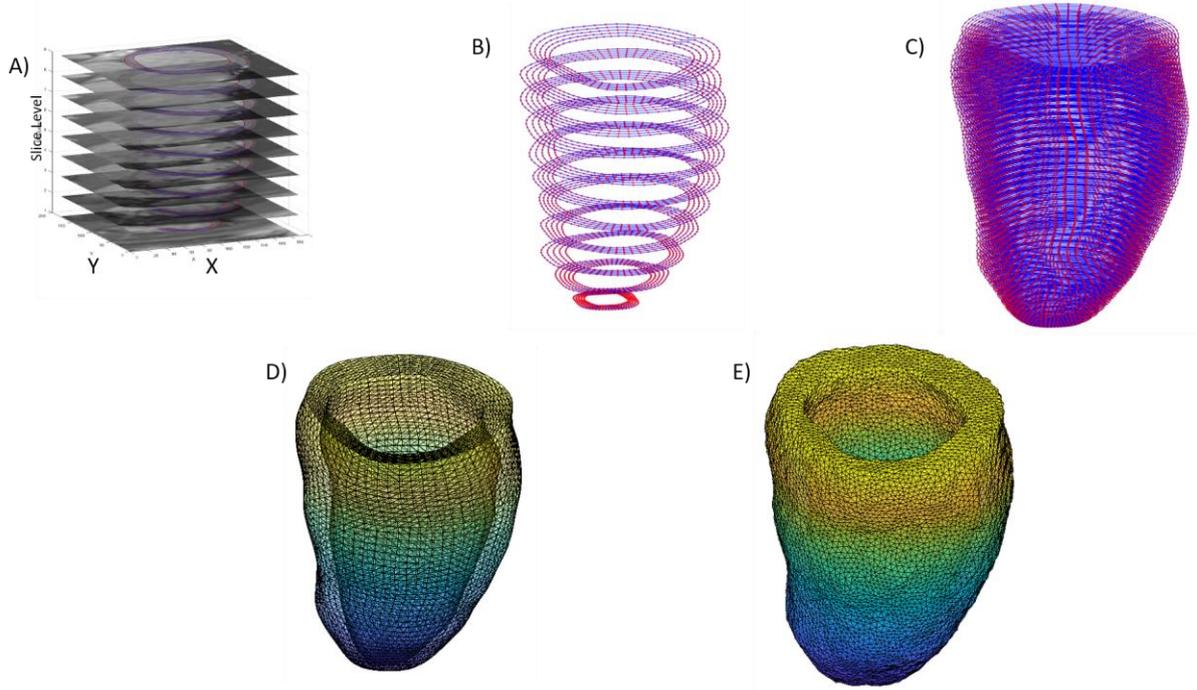

Figure 3. Overview of meshing pipeline in our proposed method. A) Initial manual segmentation of myocardium from images at end-diastole. B) Initial point cloud of tracked points from Segment. C) Up sampled and refined point cloud. D) Initial surface reconstruction. E) Final volumetric mesh of myocardium.

**Displacement Field Interpolation**

Given the aligned final mesh and originally sampled tracked pointsclouds, it is necessary to interpolate the displacement fields at arbitrary locations over time and space. Here, we rely on the scattered interpolant method [25] to compute interpolating functions for three axes of motion (X, Y, Z) by using function values from both the short axis tracked point cloud and the reconstructed long axis point cloud (Fig. 3A, B). The scattered interpolant relies on first constructing a Delaunay triangulation of the original sampled points. Next, any new query point is projected onto the convex hull of the sampled area, and the function value can be calculated using natural spline, linear or nearest neighbour interpolation options. When the query point lies inside the convex hull, the function value is interpolated, otherwise it is extrapolated using only the linear or nearest neighbour options. Finally, we can use the interpolants to compute displacement vectors at each time frame (Fig. 4). In particular, we select the Z component of the displacements at a point to be exclusively represented by the interpolated values that come from the long axis view, given that we do not track long axis motion in short axis views, while we represent the X and Y components of displacement by a time-dependent weighted combination of long (4CH, 2CH) and short axis (SAX) views (Fig. 4). The weights $W$ influence the final displacements $u$, $v$ in the X and Y directions at a point, depending on how much the given point has moved along the longitudinal direction $w_l$ at a given time frame. $W$ favours x-y displacements, represented by $u_l, v_l$ respectively, obtained from the long axis views at times of large longitudinal deformation only, according to the formula.

$$W = \frac{w_l}{(w_{l_{min}} - w_{l_{max}})} \quad (4)$$

$$u = (1 - W) * u_{cs} + W * u_l \quad (5)$$

$$v = (1 - W) * v_{cs} + W * v_l \quad (6)$$

Conversely, $u_{cs}$, $v_{cs}$ the displacement vectors for x-y directions from short-axis estimation are have a weaker influence in periods of large longitudinal deformations due to the $(1 - W)$ term.

The weight matrix $W$ is the same size as the points in the mesh assigning to each point a normalized weight between 0-1, representing the percentage of longitudinal displacement experienced by the point in a particular time frame from original position. In a way, this approach mimics enforcing incompressibility albeit in a more global coarse manner, to allow for myocrdium element deformation that adapts to the 2D estimates depending on their relative magnitudes.

Finally, with the obtained point (nodal) displacements, we compute the Greene-Lagrange strain $E$ (Eq. 3) for each tetrahedral element via the deformation gradient $(F)$ [8]. The assembly of the required matrix calculations can be derived by standard finite element mapping [26].

$$E = \frac{1}{2}(F^T F - I) \quad (7)$$

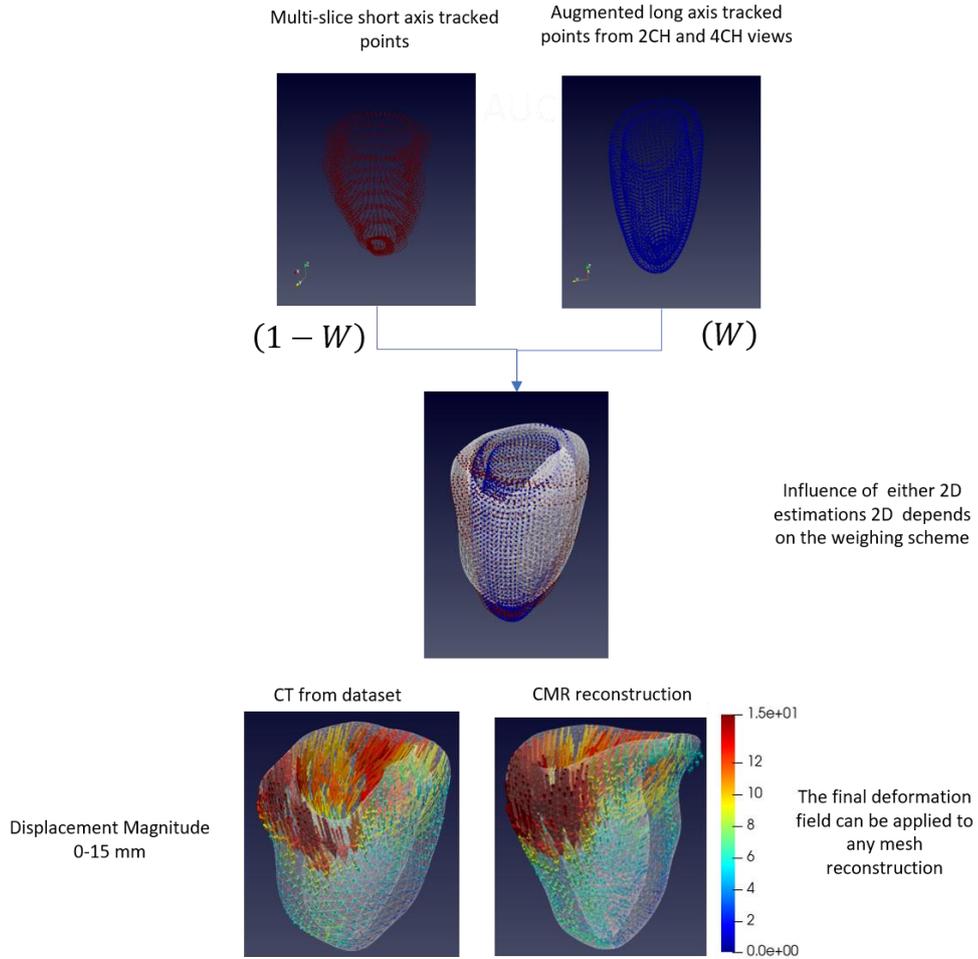

Figure 4. Overview of displacement interpolation scheme. As we show in the second panel, the both the conformed tracked points (sax, lax) encode displacement information in the X-Y directions we combine their influence via the weighing scheme. The scattered interpolant functions allow deformation of any arbitrary meshes from CT or our proposed CMR reconstruction.

**Validation and Analysis**

For validation of our procedure, we focus on LV strain estimation on SSFP images. We here select the three benchmark results that were produced on such images in the STACOM 2011 challenge [4]. Furthermore, we compare with the IUCL results also presented in STACOM 2011, which included a combined approach between 3DTag and SSFP. We begin by a qualitative assessment of the strain field maps for a single volunteer dataset as applied using the ground truth (GT) mesh representation. We inspect local morphology changes during peak systole and compare visually to patterns obtained by STACOM 2011 participants. We then discuss the effect of using different motion tracking approaches and imaging modalities on the distribution of deformation and strain. Quantitative analysis is then performed for global peak strains. Peak strains were averaged for all AHA 16 segments across all samples. Results obtained from our proposed procedure will be tagged in the results section of this paper as AUC and AUC_MESH when comparing to the results reported by challenge participants (as available) and to suitable clinical reference benchmarks [4,27,28]. Finally, we also included the 2D Segment strain results as an additional ground truth comparison. Note that literature characterizes for end-systole a positive radial strain, a negative circumferential strain, and a longitudinal strain (of lower absolute magnitude than the radial component) [27,28].

# Results

We here present our results for our procedure as applied to the STACOM 2011 datasets of the LV, which offers a valuable benchmarking platform for myocardial deformation and tracking algorithms. STACOM 2011 publicly shared its organized datasets of high quality, as well as the clinical metrics that were quantified from those datasets by four of the research groups that had responded to the challenge, the datasets are available online via the Cardiac Atlas Project [4]). Each participant proposed their own myocardial motion tracking solution for the LV based on cine SSFP, tagged MR and/or 3DUS modalities. Each human volunteer's dataset was selected from healthy volunteers within certain age limits (28 ± 5 years old) and body surface area (1.8 ± 0.2m$^2$), to reduce physiological variability[4].

**Strain Patterns in 3D**

At peak systole, the radial strain ranges from **-0.5** to **0.5**, but is predominantly positive (indicating wall thickening), with the larger values more-or-less evenly distributed along the anterior (right side, free wall) of the ventricle, and the lower values appearing towards the interior (left side, septal) region. Minimum radial contraction appears at the most basal region (held back by atrial and valvular structures) and apical region (with the least amount of muscle) (**Fig. 5**). Our results for this particular dataset, are visually comparable to those of UPF and INRIA in terms of field strength and general distribution. For the entirety of the cohort (Fig. 6) the radial strain distribution follows a similar trend but is clearly superior against IUCL and UPF.

Furthermore, the circumferential strain ranges from **-0.25** to **0.25**, and is predominantly negative (which indicates a shrinking LV chamber). Positive values, however, are visible at the most basal anterior parts, due to a loss in circularity in that plane at peak systole (**Fig. 5**). Our circumferential strain patterns for this dataset, seem to align best with the results of INRIA and UPF as well. When looking at the strain distribution for the entire cohort (Fig. 6), we see a better representation of circumferential strain for the basal and mid-ventricular regions against all participants.

The longitudinal strain correspondingly ranges from **-0.25** to **0.25** and is mostly negative over the mid-ventricular region of the LV, with positive values appearing at the most basal anterior parts. Our results are qualitatively in moderate agreement with those of UPF, **(Fig. 5)**. Finally, in terms of behaviour for the entire 15 sample cohort, we observe similar representation of longitudinal strain.

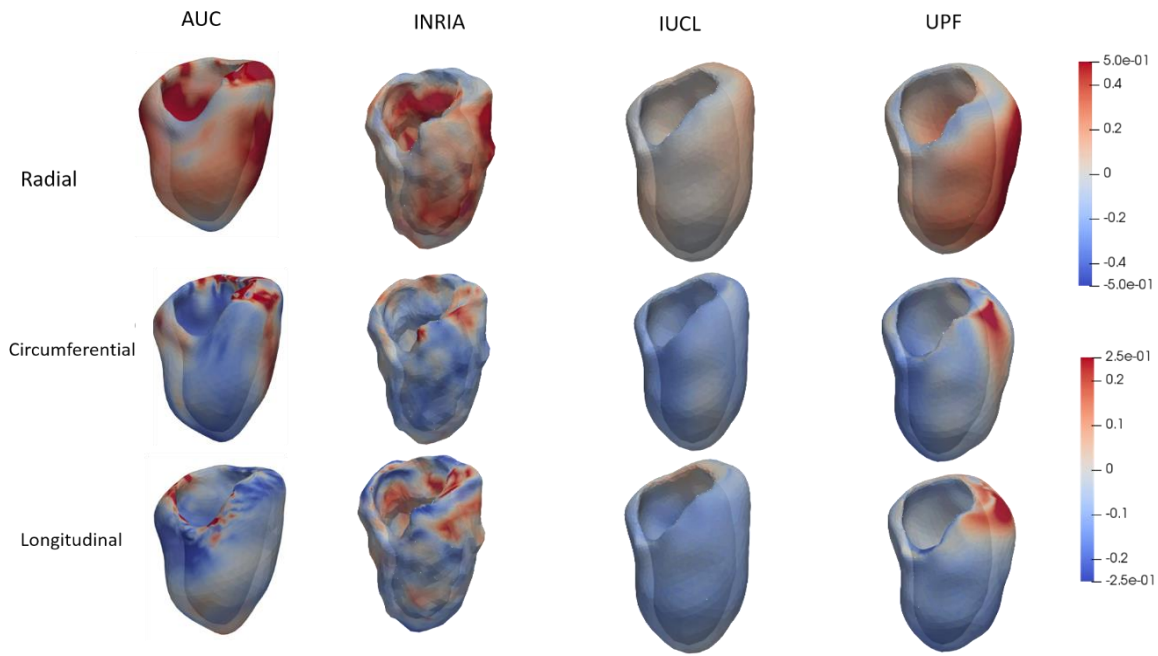

Figure 5. : Comparison of strain maps over volunteer 9 dataset between our proposed method and the results of the different participants on the same image modality. Strain ranges from –0.5-0.5 for radial, and –0.25-0.25 for longitudinal and circumferential components respectively.

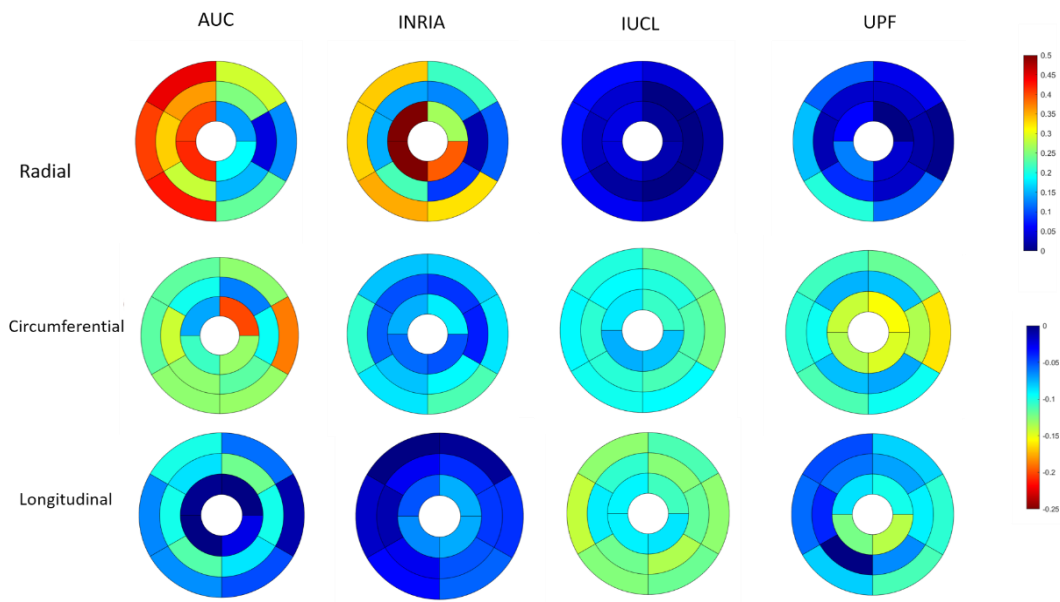

Figure 6. Comparison of strain maps over the entire dataset between our proposed method and the results of the different participants on the same image modality. Strain ranges from 0-0.5 for radial, and -0.25-0 for longitudinal and circumferential components respectively.

**Effect of Mesh Choice**

Here we present a representative figure showing the difference in deformation at peak systole for our developed CMR reconstruction and the provided ground truth CT mesh (**Fig. 7, Video. 1**). We remark here that although CMR reconstruction suffers from coarse feature representation, particularly in the most basal and apical regions. The displacement distribution is similar. However, these geometrical differences, may present themselves prominently in the individual mesh elements which will have implications on strain calculation. As we see in (**Fig. 8**), radial strain is underestimated for GT meshes compared to reconstruction for all three of the strain components.

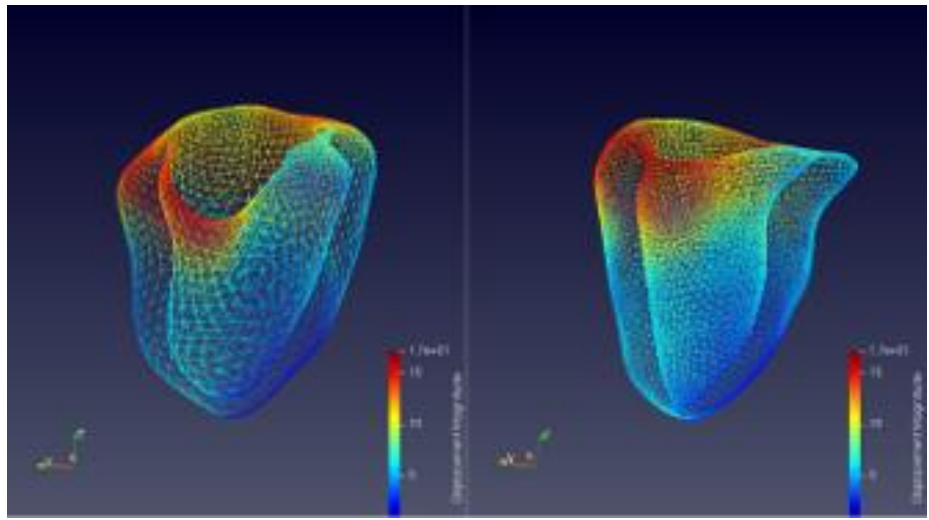

Figure 7. : Comparison of displacement and mesh geometry at peak systole for the provided ground truth CT mesh (left) and our proposed mech reconstruction (right) with displacement magnitude ranging from 0-17mm.

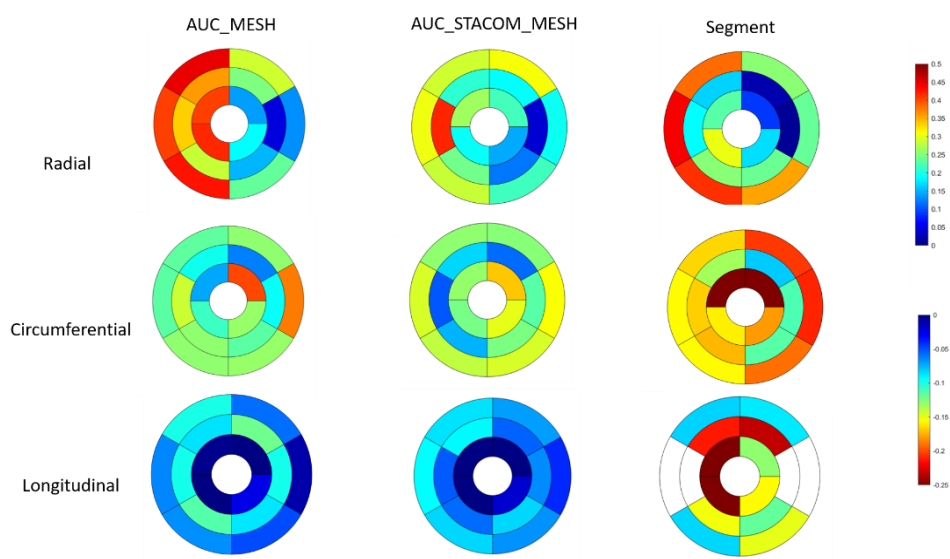

Figure 8. Comparison of strain maps over the entire dataset between our proposed method in application to GT mesh (AUC_STACOM_MESH) and 2D Segment strain results

Global **Peak Strain**

The mean global peak radial strain is estimated using our procedure to be (0.27 ± 0.13) vs. the (0.28 ± 0.15), (0.03 ± 0.15), (0.06 ± 0.14) reported by the challenge participants listed in the order shown in **(Fig. 9)**. For the circumferential strain we estimate a mean peak value of (-0.12 ± 0.04) vs. the (-0.06 ± 0.05), (-0.1 ± 0.05), (-0.12 ± 0.05) reported by the challenge participants. Finally, for the longitudinal strain is estimated to be (-0.05 ± 0.06) vs. the (-0.04 ± 0.07), (-0.11 ± 0.07), (-0.08 ± 0.06) reported by the challenge participants. The underestimation is with respect to the reference clinical benchmarks at the mid-ventricular region, as reported in the literature, i.e. (0.44 ± 0.2), (-0.19 ± 0.04) and (0.15 ± 0.03) for radial, circumferential and longitudinal strain components respectively **(Fig. 9)**, based on 3DTag imaging [27,28]. Additionally, when we look at the performance of our proposed method on either the GT mesh or our reconstruction (AUC and AUC_MESH) respectively, we notice very similar ability for circumferential and longitudinal components, with higher radial strain for the reconstruction. In terms of results with respect to 2D Segment, a more representative ground truth evaluation, we notice that all methods severely underestimated longitudinal strain, except for IUCL.

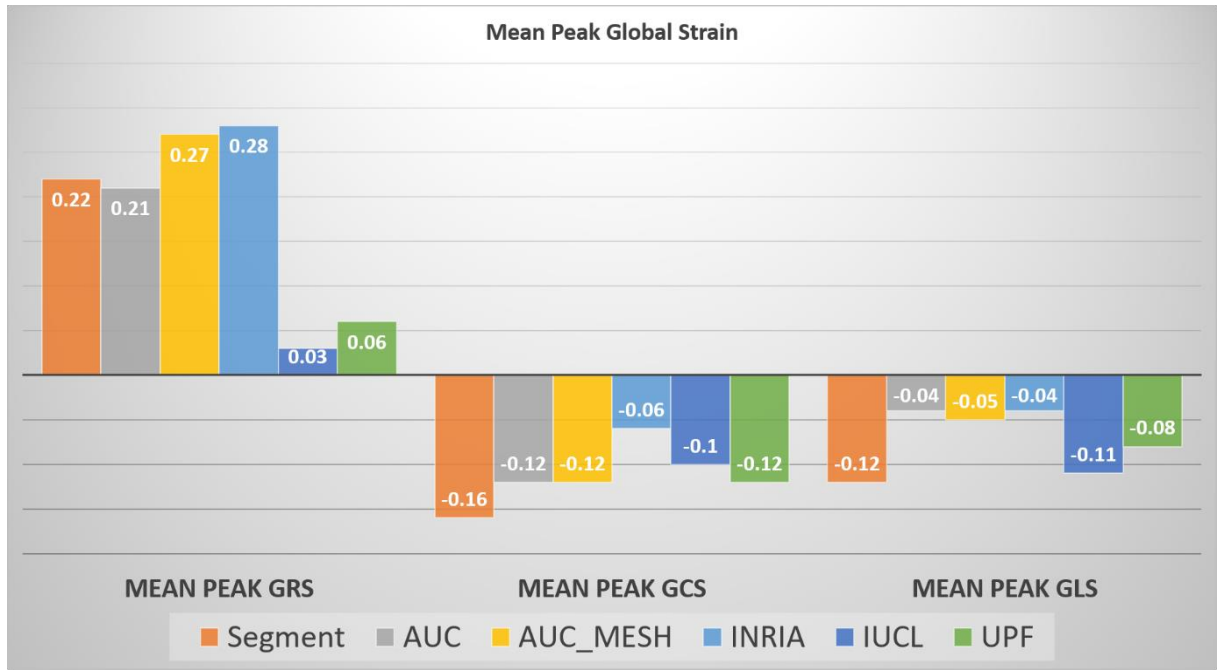

Figure 6. Bar chart comparing peak global strain across the different challenge results and our experiments with AUC, AUC_MESH and Segment respectively

## Discussion

In terms of deformation pattern in 3D **(Fig. 5)**, our model is closer in strain distribution to INRIA and UPF for both the radial and circumfriential components and closer to INRIA for the longitudinal component. Also, our deformed mesh exhibits smooth surface deformations, similar to (IUCL and UPF), with notable radial dilation in myocardium**.** Looking, at the average behaviour across the entire dataset, we note that our modelled strain maps for radial and circumferential components are concordant with the 2D Segment strain estimates **(Fig. 8)**, and in terms of global peak strains are in agreement with estimates other particpants (UPF, IUCL) but lower than 2D Segment **(Fig. 7, 8, 9)**. . We also, remark here on, the effect of mesh choice, in terms of using the provided GT CT mesh or our developed CMR mesh reconstruction (**Fig. 7, 8**). With lower spatial resolution in the long axis view, the mesh reconstruction is particularly weak in capturing shape features towards the basal and apical parts of the heart. Which, possibly affects strain estimations based on finite element approximations, even if displacement patterns are equivalent. As the LV walls thicken between diastole and systole within a cardiac cycle, with an underestimated longitudinal motion, an enlarged myocardial volume will result, which violates incompressibility. All computed strain components are thus calculated with respect to enlarged tetrahedral elements, which for the same displacements will return underestimated strains

In regards to improvements offered by our procedure, with respect to the results reported in the challenge. We observe, smooth deformations that produce global peak strains, closer to reference values in literature [4,27,28] as well as higher sensitivity in capturing peak strain patterns in the different directions. Second, in terms of general mesh smoothness [4,16] which may indicate that our method is more robust to noisy acquisitions and myocardial tracking errors. Concerning our modelled strain magnitudes, as with all challenge participants, we underestimated the peak value, with varying degrees for each component. Here, we compute a peak radial strain of ($27 \pm 13$ %) (**Fig. 7, 9**), consistent with one of the challenge participants and with previous studies made on SSFP images only [4,16]. We also have good agreement in circumferential strain ($12 \pm 4$%) with the participants (**Fig. 5, 9**), when focusing on the SSFP modality. Our approach however fails to reproduce the higher longitudinal strain measurements reported by other modalities (3DTAG or 3DUS) . Similar, to all other particpants, except for IUCL group, employing their multi-modal approach on 3DTAG and SSFP, were able to capture at the expense of radial component (**Fig. 6, 9**). One possible reason is the lack of a parasternal long axis view (3CH view) in the chosen dataset, with our long axis motion estimates dependent on partial coverage of myocardium, combined with our interpolation scheme, leads to weaker representation of long axis motion in missing parts of the myocardium which explains the underestimation even compared to the 2D results (**Fig. 8, 9**), particularly for the basal and apical sectors.

Furthtermore, we remark that another source of underestimation affecting strain estimates in all the approaches based solely on SSFP images, is the inherent low resolution in long axis information with only 2 slices (4CH, 2CH) compared to up to 16 for the short axis. Long axis estimations, will always be subject to overestimation in 2D due to possible emergence of myocardium tissue in the long axis image plane due to torsion.

Only, one participant enforced incompressiblity in their approach (INRIA), however, we note the notable nosiy deformations (**Fig. 6**), which we hypothesise is due to local incompressiblity constraints, forcing tissue elements to "bulge" or "shrink", to maintain the constraint [7–9,16]. And, still being unable to deform sufficiently in the long axis direction.

We finally, note that we were able to generate significantly coarser meshes to obtain comparable results to those of the participants with the ground truth CT mesh (**Fig. 7, 8, 9**), which may provide reasonable accuracy in the common scenario of lack of CT imaging for a particular patient, due to diagionistic guidelines or presence of pacemakers or other artificial implants.

## Limitations

Our modelled approach, as evidenced by our comparisons on effect of mesh choice. Is particularly sensitive to initial alignment errors between, the initial tracked points in 2D and the subject of deformation. Hence, an accumulation of errors is possible due to either operator contouring error or disagreement in location of local shape features between CMR and mesh designed using a different image modality such as CT. As the scattered interpolation scheme, will have to rely on data extrapolation rather than interpolation, when, parts of the myocardium are outside the field of view of the displacement fields. Furthermore, the apparent underestimation in the longitudinal strain, is possibly due to the combined effect of misalignment in the myocardium edges (either the basal or apical regions) as well as lack of enough data by relying on SSFP images only. Finally, although our weighing scheme was designed to counteract overestimation of in plane strain components in 3D, as a simpler implementation to explicit incompressiblity. It was only successfully in generating smoother deformation patterns, but not in achieving more accurate longitudinal shortening. This is partly due to the lack of the 3$^{rd}$ missing 3CH view as explained previously.

## Conclusions

Motion analysis of a beating heart has important clinical implications in the characterization of heart disease. It is associated with structural and functional changes with respect to normal conditions. Rapid advancements to imaging technologies, image processing, image registration and computer visualization afford us an opportunity for patient-specific quantification of normal and abnormal patterns of deformation. These advancements promise a better understanding of the underlying cardiac conditions to support standard clinical practices, such as diagnosis, pre-operative surgery planning, and post-operative assessment. In this work we have set out to develop a computational framework that generates a four-dimensional model of the LV, which incorporates information on LV structure and deformation through geometric approaches that interpolate between a set of 2D displacements, using 3D spatial interpolation. We developed an end-to-end procedure for motion analysis by generating our 3D geometric representation from the 2D data, and by creating a volumetric mesh representation for the myocardium, which allows a more accurate representation of myocardial deformation (via finite element approximation functions that capture through-thickness strain components). Our procedure is computationally inexpensive and fully automated, except for the initial segmentation step. Our procedure also does not rely on multi-modal datasets, special imaging methods or additional image annotation, only the commonest imaging modality with standard LV contouring in the first time frame based on tracking results provided by one of the few clinically used software for LV function assessment on untagged CMR images, to produce fast quality predictions of strain (radial, circumferential and longitudinal), which is a key feature sought by our procedure for broad clinical and patient-specific application. And can be readily integrated into existing clincal workflows prescribed for LV function assessment on CMR images.

## Data Availability

The unprocessed data sets used in this research are publicly available on the cardiac atlas project as part of the STACOM 2011 challenge dataset. The processed vtk files which contain the resulting data from applying the framework which was used to support the findings of this study are available from the corresponding author upon request.

## Conflicts of Interest

The authors declare that they have no conflict of interest.

## Funding Statement

This research did not receive any specific grant from funding agencies in the public, commercial, or not-for-profit sectors.

## Supplementary Materials

1) Video animation of comparing deformation over the cardiac cycle of the, final interpolated mesh (CT vs CMR) as it deforms due to influence of combined displacement from the multiple views using our developed method